\documentstyle{article} \textheight=670pt \textwidth=450pt
\begin{document}

\title{\Large {\bf Revisiting controlled quantum secure direct communication using a non-symmetric
quantum channel with quantum superdense coding}}
\author{Jun Liu$^1$, Yan Xia$^3$ and Zhan-jun Zhang$^{1,2,} \thanks{Corresponding author. Email: zjzhang@ahu.edu.cn}$ \\
{\normalsize $^1$ Key Laboratory of Optoelectronic Information
Acquisition \& Manipulation of Ministry of Education of China,}\\
{\normalsize School of Physics \& Material Science, Anhui University, Hefei 230039, China} \\
{\normalsize $^2$ Department of Physics and Center for Quantum Information Since,}\\
{\normalsize National Cheng Kung University, Tainan 70101, Taiwan}\\
{\normalsize $^3$School of Physic and Optoelectronic, Dalian
University of Technology, Dalian 116024, China} }

\date{\today}
\maketitle

\begin{minipage}{420pt}

{\bf Abstract} Recently Xia and Song [Phys. Lett. A (In press)] have
proposed a controlled quantum secure direct communication (CQSDC)
protocol. They claimed that in their protocol only with the help of
the controller Charlie, the receiver Alice can successfully extract
the secret message from the sender Bob. In this letter, first we
will show that within their protocol the controller Charlie's role
can be excluded due to their unreasonable design. We then revise the
Xia-Song CQSDC protocol such that its original advantages are
retained and the CQSDC can be really realized.\\



\end{minipage}\\

Different from quantum key distribution (QKD)[1-2] whose object is
to distribute a common key between the two remote legitimate users
of communication, quantum secure direct communication (QSDC) can
transit the secret messages directly without creating first a key to
encrypt them. Lately, many QSDC protocols have been proposed and
actively pursued by some groups[3-20] since Beige et al.[3] first
proposed a QSDC protocol taking advantage of Einstein-Podlosky-Rosen
(EPR) pairs. Very recently, Xia and Song[10] presented a controlled
QSDC protocol using a non-symmetric quantum channel with quantum
superdense coding (referred to as the Xia-Song CQSDC protocol
hereafter).

Xia and Song [10] claimed that their protocol is a novel protocol
for the implementation of controlled QSDC by using the particles in
different dimension's Hilbert space. In their protocol they combined
the ideas of block transmission, entanglement swapping, and quantum
dense coding in the non-symmetric quantum channel. The communication
is secure under some eavesdropping attack and two users can transmit
their secret messages with the help of controller Charlie securely
and simultaneously. The efficiency of information transmission is
also successfully increased. We agree that all these are advantages
of their protocol except for one very important point. Since the
Xia-Song CQSDC protocol is a controlled QSDC protocol, the
controller's role should be indispensable. Nevertheless, we find
that in their protocol the so-called controller Charlie is unvalued
in fact. In this letter we will point out this leak and explain why
Charlie is unvalued. Moreover, we will revise the Xia-Song CQSDC
protocol such that its original advantages are retained and the
CQSDC can be really implemented.

Before pointing out the leak in the Xia-Song CQSDC protocol, let us
briefly review the scenario Xia and Song considered and their two
design ideas which we think are quite unreasonable. Xia and Song
consider the following scenario. Suppose there are three involved
parties Alice, Bob and Charlie. Alice and Bob are the secret message
receiver and sender, respectively. Alice and Bob want to implement
their direct secret communication. Charlie is a controller
introduced to control the secret communication between Alice and
Bob. To achieve this goal, when designing their protocol, Xia and
Song take the following two ideas. (1) Alice is assigned to prepare
a series ($N$) of 2-dimensional GHZ entangled states in
$|\Psi_1\rangle_{ABc}={\frac{1}{\sqrt{2}}}(|000\rangle+|111\rangle)_{ABc}$
and a series ($M$) of 3-dimensional Bell-basis states in
$|\Phi\rangle_{A'B'}={\frac{1}{\sqrt{3}}}(|00\rangle+|11\rangle+|22\rangle)_{A'B'}$.
She transmits the $S_B$ sequence (consists of all $B$'s and $B'$'s)
to Bob for later secret message communication and the $S_c$
(consists of all $c$'s) sequence to Charlie as his later control.
That is to say, Alice is the quantum channel constructor. (2) Alice
is assigned to finally judge whether the quantum channel is secure
through comparing her measurement results with Bob's and Charlie's.
If Alice announces that the quantum channel is secure, then Bob
sends his secret message via his encodings and thinks Alice can
extract the secret message only with Charlie's help. That is to say,
Alice is also the quantum channel security checker.

Now let us extensively analyze two design ideas in the Xia-Song
CQSDC protocol. It is intriguing to ask, in the the scenario Xia and
Song considered, who wants to let the QSDC be controlled by the
controller Charlie? Essentially, this is an very important question
and should be in prior considered. Otherwise, an unreasonable
protocol may be designed, as can be seen in the Xia-Song CQSDC
protocol. In the Xia-Song CQSDC protocol, Alice is the secret
message receiver. Is she willing to receive Bob's secret message
with control of Charlie? Of course, in a normal case, she is not.
Then why she sends the $S_c$ sequence to Charlie as his control of
the communication between her and Bob, can only be explained as the
demand of Bob (not of Charlie, because Alice is the quantum channel
constructor. If it is Charlie who demands Alice to let him control
the secret communication between Alice and Bob, Alice can disregard
Charlie's demand or just sends him a fake signal to cheat him. In
this case, the communication between her and Bob can be successfully
implemented). That is to say, Bob wants to let Alice retrieve his
secret message only with Charlie's help. In this case, if Alice
wants to exclude Charlie's control, what can she do? She can send a
fake sequence $S_c'$ to Charlie and keeps the sequence $S_c$ in her
site. If so, obviously both Bob and Charlie can not find Alice's
this cheating, because in the Xia-Song CQSDC protocol Alice is the
party who finally judges whether the quantum channel is secure.
Consequently, after Bob encodes his secret message, Alice can obtain
Bob's secret message without Charlie's help. This is the leak of the
Xia-Song CQSDC protocol. The controller's role can be excluded by
the receiver. The basic reason which induces this unexpected result
is that Alice is assigned to be not only the quantum channel
constructor but also the quantum channel security checker.

If let different parties to act as the quantum channel constructor
and the quantum channel security checker, respectively, then it
seems that the leak in the Xia-Song CQSDC protocol can be fixed
immediately. In fact, if Alice is still assigned as the quantum
channel constructor or the quantum channel security checker, it is
quite possible that she has potential to cheat the other two such
that she can exclude Charlie's control and obtain Bob's secret
message without Charlie's help. Because the secret message is
initially in Bob's hand, only if he thinks that the quantum channel
is secure he will encode his secret message. Otherwise, the CQSDC is
aborted. Hence in the following we will revise the Xia-Song CQSDC
protocol in such a way that Bob is assigned to be both the quantum
channel constructor and the quantum channel security checker, its
original advantages are retained and the CQSDC can be really
realized.  Moreover, in the revised version the security check of
quantum channel is varied in terms of randomly choosing measuring
basis so that the security of quantum channel can be assured. One
will see this later. Hence the revised version is not a simple
exchange between Alice and Bob of the original Xia-Song CQSDC
protocol. For convenience, in the revised version we will quote some
descriptions and notations in the original Xia-Song CQSDC protocol.
For completeness, we detailedly show the 9 steps of the revised
version as follows:

(S1) Bob prepares a series ($N$) of 2-dimensional GHZ entangled
states in
\begin{equation}
|\Psi_1\rangle_{ABc}={\frac{1}{\sqrt{2}}}(|000\rangle+|111\rangle)_{ABc} ,
\end{equation}
and a series ($M$) of 3-dimensional Bell-basis states in
\begin{equation}
|\Phi\rangle_{A'B'}={\frac{1}{\sqrt{3}}}(|00\rangle+|11\rangle+|22\rangle)_{A'B'},
\end{equation}
where particles $A$, $B$ and $c$ are all in 2-dimensional Hilbert
space, particles $A'$ and $B'$ are both in 3-dimensional Hilbert
space, and $A$ ($A'$) represents {\it "travel"}, $B$ ($B'$) labels
{\it "home"}, and $c$ stands for {\it "control"}. Here Alice and Bob
agree on that the six collective unitary operations $U_1$, $U_2$,
$U_3$, $U_4$, $U_5$ and $U_6$ (see Eq. (9) in Ref.[10]) correspond
to the secret messages 00, 01, 10, 11, 20 and 21, respectively.
During the security check, they can use two sets of measuring bases
(MBs), $Z$$\equiv$$\{|0\rangle,|1\rangle\}$ and
$X$$\equiv$$\{|+x\rangle={\frac{1}{\sqrt{2}}}(|0\rangle+|1\rangle),
|-x\rangle={\frac{1}{\sqrt{2}}}(|0\rangle-|1\rangle)\}$ to measure
the sample particles of 2-dimensional GHZ entangled state randomly.
For the 3-dimensional Bell-basis state, there are four such complete
bases. The Z-MB is composed of the eigenvectors
$|Z_{-1}\rangle=|0\rangle$, $|Z_{0}\rangle=|1\rangle$ and
$|Z_{+1}\rangle=|2\rangle$. The X-MB is chosen as
\begin{eqnarray}
&&|X_{-1}\rangle={\frac{1}{\sqrt{3}}}(|0\rangle+|1\rangle+|2\rangle),
 \\
&&|X_{0}\rangle={\frac{1}{\sqrt{3}}}(|0\rangle+e^{2\pi{\it
i}/{3}}|1\rangle+e^{-2\pi{\it i}/{3}}|2\rangle), \\
&&|X_{+1}\rangle={\frac{1}{\sqrt{3}}}(|0\rangle+e^{-2\pi{\it
i}/{3}}|1\rangle+e^{2\pi{\it i}/{3}}|2\rangle).
\end{eqnarray}
The two other bases are formed by
\begin{equation}
{\frac{1}{\sqrt{3}}}(e^{2\pi{\it
i}/{3}}|0\rangle+|1\rangle+|2\rangle) {\rm \ \ and \  cyclic  \
permutation}
\end{equation}
 and
\begin{equation}
{\frac{1}{\sqrt{3}}}(e^{-2\pi{\it
i}/{3}}|0\rangle+|1\rangle+|2\rangle) {\rm \ \ and \  cyclic  \
permutation}.
\end{equation}
Alice and Bob randomly use one of the four MBs to measure the sample
particles of 3-dimensional Bell-basis states during the security
check.

Bob takes particle $A'$ from each 3-dimensional Bell-basis state and
particle $A$ from each 2-dimensional GHZ entangled state to form an
ordered particle groups $[P_1(A'A), P_2(A'A), . . . ,P_N(A'A)]$,
called $S_A$ sequence. Then he sends the $S_A$ sequence to Alice.
Where the subscripts indicate each 2-dimensional GHZ entangled
state's and each 3-dimensional Bell-basis state's order in the
sequence. Similarly he generates $S_B$ sequence and keeps it in his
site. He transmits the particles $c$ (called the $S_c$ sequence
hereafter) to Charlie in order.

(S2) Alice and Charlie publicly confirm that they have received the
$S_A$ sequence and $S_c$ sequence, respectively. They store the
$S_A$ sequence and $S_c$ sequence according to their coming orders.
If Bob wants to transmit messages to Alice, he randomly picks out a
sufficiently large subset of particle groups from the $S_B$ sequence
as sample particle groups to check the security of the transmission.
The remaining particle groups in the $S_B$ sequence are taken as
encoding-decoding particle groups for his later encoding via local
unitary operations $U_m$ $(m\in{1, 2, . . . , 6})$.

(S3) For each checking group, the security check consists of two
parts, i.e., the security check of 2-dimensional GHZ entangled state
and the security check of 3-dimensional Bell-basis states, which can
be completed with the following procedures. (a) Bob tells Alice and
Charlie the sample particle groups that he has chosen. Then Alice
and Charlie pick out the corresponding particle groups and particles
from $S_A$ sequence and $S_c$ sequence. For instance, if Bob chooses
$P_1(B'B)$, Alice and Charlie should choose $P_1(A'A)$ and $c_1$.
(b) Bob randomly chooses the Z-MB or X-MB to measure the chosen
particles $B'$ and $B$. (c) Alice and Charlie also randomly choose
the Z-MB or X-MB to measure the corresponding particles $A'$, $A$
and $c$. (d) Bob randomly selects Alice or Charlie to tell him the
corresponding MBs chosen and the measurement results. Such random
ordering of who tell first can prevent cheating. This check
procedure is an important improvement of the original protocol. (e)
Alice, Bob and Charlie have 25\% probability to choose the same MBs,
i.e., they all choose Z-MB or X-MB simultaneously, when they all
choose the same MBs, Bob compares his measurement results with
Alice's and Charlie's to check the existence of Eve. The
intercept-and-resend attack as well as the entangle-and-measure
attack can be detected efficiently by this method. If no
eavesdropping exists, their measurement results should be completely
correlated in an ideal condition. For instance, during the security
check of 2-dimensional GHZ entangled state, if their measurement
outcomes coincide when they use the same basis in order according to
Eqs. (11)-(18) in Ref.[10], they go to check 3-dimensional
Bell-basis states' security. Otherwise, they have to discard their
transmission and abort the communication. If there is no Eve in
line, the procedure goes to (S4), otherwise they discard the
communication.

(S4) Bob asks Charlie to perform a Hadamard operation on each $c_n$
in order. The Hadamard operation takes the form
\begin{equation}
H|0\rangle={\frac{1}{\sqrt{2}}}(|0\rangle+|1\rangle),
\end{equation}
\begin{equation}
H|1\rangle={\frac{1}{\sqrt{2}}}(|0\rangle-|1\rangle),
\end{equation}
which will transform the state shown in Eq. (1) into
\begin{equation}
|\Psi_1\rangle_{n}={\frac{1}{\sqrt{2}}}[(|00\rangle+|11\rangle)_{A_nB_n}\otimes|0\rangle_{c_n}
+(|00\rangle-|11\rangle)_{A_nB_n}\otimes|1\rangle_{c_n}].
\end{equation}
If Charlie would like to help Alice and Bob to communication, he
first performs a Hadamard operation on each $c_n$ in order.  Then he
measures photon $c_n$ and informs Alice and Bob of his measurement
results via classical communication. The procedure goes to (S5);
otherwise, Charlie do nothing on photon $c_n$. Without the help of
controller Charlie, there will be no perfect quantum channel between
Alice and Bob. As a consequence, Alice cannot extract Bob's secret
messages solely (she can only read 1 bit information), and  the
controlled quantum secure direct communication protocol failed.

(S5) Charlie measures his photon $c_n$. If Charlie obtains the
outcome $|0\rangle_{c_n}$, photons $(h_n, t_n)$ will collapse into
the state
\begin{equation}
|\Phi_1^+\rangle_n={\frac{1}{2}}(|00\rangle+|11\rangle)_{A_nB_n};
\end{equation}
otherwise, we propose that the state of photons $(h_n, t_n)$ will be
\begin{equation}
|\Phi_1^-\rangle_n={\frac{1}{2}}(|00\rangle-|11\rangle)_{A_nB_n}.
\end{equation}
After the above operations, in accord with the encoding-decoding
group ordering, Charlie informs Alice and Bob of his measurement
results via classical communication.

(S6) In accord with the encoding-decoding particle groups ordering,
Bob encodes his secret messages by performing one of the six local
collective unitary operations $U_1$, $U_2$, $U_3$, $U_4$, $U_5$ and
$U_6$ on particles $B'$, $B$ of each $P_i(B'B)$ except the sample
particles according to his secret messages (say, 0001 . . .) need to
be transmitted this time: for instance, $U_1$ operation on group 1
to encoding 00. $U_2$ operation on group 2 to encode 01, etc. After
the unitary $U$ operations, Bob makes the measurements on particle
groups $B'$, $B$ under the basis \{$|\Psi_{00}\rangle$,
$|\Psi_{01}\rangle$, $|\Psi_{10}\rangle$, $|\Psi_{11}\rangle$,
$|\Psi_{20}\rangle$, $|\Psi_{21}\rangle$\} (see Eq. (3)-(8) in
Ref.[10]).

(S7) Bob publicly announces his measurement results and the position
of the particle group $B'$ and $B$.

(S8) Alice measures the particle groups $A'$, $A$ of the
corresponding $P_i(A'A)$ except the sample particles under the basis
\{$|\Psi_{00}\rangle$, $|\Psi_{01}\rangle$, $|\Psi_{10}\rangle$,
$|\Psi_{11}\rangle$, $|\Psi_{20}\rangle$, $|\Psi_{21}\rangle$\}.
After the measurements, Alice can conclude the exact local
collective unitary operations performed by Bob according to Bob's
measurement results and the position of particle groups $B'$, $B$
and her own measurement results of particles $A'$, $A$. Hence Alice
can successfully get Bob's secret messages.

(S9) The controlled QSDC has been successfully completed.

So far, we have shown the revised version of the Xia-Song CQSDC
protocol. If the sender Bob wants to transmit messages to the
receiver Alice, the controller Charlie must take a Hadamard
operation on each of his particles, measure it, and tell the results
to Bob and Alice in order. Our version can also be generalized to
multi-party control system. $N$-party share a large number of
$N$-particle 2-dimensional GHZ entangled states. Party $M$ $(M\in
N)$ as receiver, party $Q$ $(Q\in N, Q\neq M)$ as sender, and $N-2$
parties except for party $M$ and party $Q$ as controller. The
multi-party controlled QSDC can also succeed. The security analyses
of our this CQSDC protocol is the same as the Xia-Song CQSDC
protocol.

To summarize, in this letter we have pointed out a flaw in the
original Xia-Song CQSDC protocol, that is, within their protocol the
controller Charlie's role can be excluded. We have revised the
original Xia-Song CQSDC protocol such that its advantages are
retained and the CQSDC can be really achieved.\\

\noindent {\bf Acknowledgements}

This work is supported by the National Natural Science Foundation of
China under Grant Nos.60677001 and 10304022, the science-technology
fund of Anhui province for outstanding youth under Grant
No.06042087, the general fund of the educational committee of Anhui
province under Grant No.2006KJ260B, and the key fund of the ministry
of education of China under Grant No.206063. \\

\noindent {\bf References}

\noindent[1] C. H. Bennett, G. Brassard, in: Proceedings of the IEEE
International Conference on Computers, Systems and Signal
Processing, Bangalore, India, IEEE, New York, 1984, p.175.

\noindent[2] N. Gisin, G. Ribordy, W. Tittel, H. Zbinden, Rev. Mod.
Phys. 74 (2002) 145.

\noindent[3] A. Beige, B. G. Englert, C. Kurtsiefer, H. Weinfurter,
Acta Phys. Pol. A 101 (2002) 357.

\noindent[4] K. Bostr\"{o}m, T. Felbinger, Phys. Rev. Lett. 89
(2002) 187902.

\noindent[5] F. G. Deng, G. L. Long, X.S. Liu, Phys. Rev. A 68
(2003) 042317.

\noindent[6] F. G Deng, G. L. Long, Phys. Rev. A 69 (2004) 052319.

\noindent[7] M. Lucamarimi, S. Mancini, Phys. Rev. Lett. 94 (2005)
140501.

\noindent[8] Z. X. Man, Z. J. Zhang, Y. Li, Chin. Phys. Lett. 22
(2005) 18.

\noindent[9] H. J. Cao, H. S. Song, Chin. Phys. Lett. 23 (2006) 290.

\noindent[10] Y. Xia, H. S. Song, Phys. Lett. A (In press),
doi:10.1016/j.physleta.2006.11.080

\noindent[11] Z. J. Zhang, Z. X. Man, Y. Li, Int. J. Quantum Inform.
2 (2005) 521.

\noindent[12] F. L. Yan, X. Q. Zhang, Eur. Phys. J. B 41 (2004) 75.

\noindent[13] T. Gao, F. L. Yan, Z. X. Wang, J. Phys. A 38 (2005)
5761.

\noindent[14] C. Wang, F. G. Deng, G. L. Long, Opt. Commun. 253
(2005) 15.

\noindent[15] A. W\'{o}jcik, Phys. Rev. Lett. 90 (2003) 157901.

\noindent[16] Z. J. Zhang, Z. X. Man, Y. Li, Phys. Lett. A 333
(2004) 46.

\noindent[17] Z. J. Zhang, Y. Li, Z. X. Man, Phys. Lett. A 341
(2005) 385.

\noindent[18] J. Liu, Y. M. Liu,  H. J. Cao, S. H. Shi, Z. J. Zhang,
Chin. Phys. Lett. 23 (2006) 2652.

\noindent[19] Hwayean Lee, Jongin Lim, HyungJin Yang, Phys. Rev. A
73 (2006) 042305 .

\noindent[20] Z. J. Zhang, J. Liu, D. Wang, S. H. Shi, Phys. Rev. A
(In press).

\enddocument